\documentclass[12pt]{article}
\usepackage{graphicx}

\input tcilatex
\begin{document}

\baselineskip 0.1667in

\begin{center}
{\large \textbf{An Ignored Mechanism}}

{\large \textbf{for the Longitudinal Recoil Force in Railguns and}}

{\large \textbf{Revitalization of the Riemann Force Law}}

\vspace{1cm}

\textsf{Ching-Chuan Su}

Department of Electrical Engineering

National Tsinghua University

Hsinchu, Taiwan

\vspace{1cm}
\end{center}

\noindent \textbf{Abstract}\textit{\ }-- The electric induction force due to
a time-varying current is used to account for the longitudinal recoil force
exerted on the rails of railgun accelerators. As observed in the
experiments, this induction force is longitudinal to the rails and can be
the strongest at the heads of the rails. Besides, for the force due to a
closed circuit, it is shown that the Riemann force law, which is based on a
potential energy depending on a relative speed and is in accord with
Newton's law of action and reaction, can reduce to the Lorentz force law.

\vspace{0.5cm}


$\vspace{1.5cm}$

\noindent {\large \textbf{1. Introduction}}\\[0.2cm]
It is known that a railgun utilizes the magnetic force to accelerate an
armature to move along two parallel rails on which it is placed. Further, it
has been reported that a recoil force, which is longitudinal to the rails
and is exerted on them, was observed during the acceleration of the armature 
\cite{Graneau87}. Based on the Biot-Savart (Grassmann) force law, the
magnetic force exerted on a wire segment of directed length $d\mathbf{l}_{1}$
and carrying a current $I_{1}$ due to another current element $I_{2}d\mathbf{%
l}_{2}$ is given by 
$$
\mathbf{F}=-\frac{\mu _{0}}{4\pi }I_{1}I_{2}\frac{1}{R^{2}}\left[ \hat{R}(d%
\mathbf{l}_{1}\cdot d\mathbf{l}_{2})-(d\mathbf{l}_{1}\cdot \hat{R})d\mathbf{l%
}_{2}\right] ,\eqno
(1) 
$$
where $\hat{R}$ is a unit vector pointing from element 2 to element 1 and $R$
is the separation distance between them. By using a vector identity it is
readily seen that the magnetic force is always perpendicular to the wire
segment carrying the current $I_{1}$. Thus the longitudinal force cannot be
accounted for by the Biot-Savart force law.

Alternatively, in some experiments the Amp\`{e}re force law 
$$
\mathbf{F}=-\frac{\mu _{0}}{4\pi }I_{1}I_{2}\frac{\hat{R}}{R^{2}}\left[ 2(d%
\mathbf{l}_{1}\cdot d\mathbf{l}_{2})-3(d\mathbf{l}_{1}\cdot \hat{R})(d%
\mathbf{l}_{2}\cdot \hat{R})\right] \eqno
(2) 
$$
is applied to account for this longitudinal recoil force \cite{Graneau87},
though this force law is not well accepted. From this law it seems that the
longitudinal force can be expected. However, it can be shown that the force
predicted from the Amp\`{e}re law is identical to the one from the
Biot-Savart law, when the force is due to a closed circuit with uniform
current as it is ordinarily. Such an identity has also been proved by two
elegant but similar approaches by using vector identities \cite{Jolly,
Ternan}, where the current is given by a volume density as it is actually
and the singularity problem which occurs when the distance $R$ becomes zero
for the self-action term is then avoided. In these derivations the
magnetostatic condition, under which the divergence of the current density
is zero, is assumed. A closed circuit with uniform current is a common case
of this condition. Some specific analytical or numerical integrations with
volume or even surface current densities \cite{Moyssides, Assis96} also
support the identity. Thereby, without doubt, the Amp\`{e}re law is
identical to the Biot-Savart law for the force due to closed circuits and
hence the longitudinal recoil force can be accounted for by neither of them.
In spite of these theoretical arguments, there remains controversy over the
experimental observations of the railgun longitudinal force and the
experimental demonstrations for the validity of the force laws [6--11].

In this investigation, it is pointed out that the railgun longitudinal force
can be accounted for by the electric induction force which as well as the
Biot-Savart magnetic force is incorporated in the Lorentz force law. This
induction force is due to a time-varying current and its direction is
longitudinal to the current. This force is of the same order of magnitude of
the magnetic force, but it appears to be ignored in the literature dealing
with railguns. As to the Amp\`{e}re force law, it has an appealing feature
that it is obviously in accord with Newton's third law of motion. This is a
consequence of the situation that the Weber force law and hence the
Amp\`{e}re force law can be derived from a potential energy of which the
involved velocity is a relative velocity between two associated charged
particles. In section 5 it is shown that the Riemann force law, which is
derived from a potential energy where the involved velocity is also
relative, can reduce to the Lorentz force law. Thus the longitudinal rail
recoil force can be accounted for by a force law which is in accord both
with the nowadays standard theory and with Newton's law of action and
reaction.

\vspace{1cm}

\noindent {\large \textbf{2. Electric Induction Force in Railguns}}\\[0.2cm]
It is well known that in the presence of electric and magnetic fields, the
electromagnetic force exerted on a particle of charge $q$ and velocity $%
\mathbf{v}$ is given by the Lorentz force law 
$$
\mathbf{F}=q\left( \mathbf{E}+\mathbf{v}\times \mathbf{B}\right) .\eqno
(3) 
$$
This force law and Maxwell's equations form the fundamental equations
adopted by Lorentz in the early development of electromagnetics. The Lorentz
force law can be given directly in terms of the scalar and the vector
potential originating from the charge and the current density, respectively.
That is, 
$$
\mathbf{F}=q\left( -\nabla \Phi -\frac{\partial \mathbf{A}}{\partial t}+%
\mathbf{v}\times \nabla \times \mathbf{A}\right) ,\eqno
(4) 
$$
where $\Phi $ is the electric scalar potential and $\mathbf{A}$ is the
magnetic vector potential. The term associated with the gradient of the
scalar potential, with the time derivative of the vector potential, and the
one with the particle velocity are known as the electrostatic force, the
electric induction force, and the magnetic force, respectively.
Quantitatively, the scalar and the vector potential are given explicitly in
terms of the charge density $\rho $ and the current density $\mathbf{J}$
respectively by the volume integrals 
$$
\Phi (\mathbf{r},t)=\frac{1}{4\pi \epsilon _{0}}\int \frac{\rho (\mathbf{r}%
^{\prime },t)}{R}dv^{\prime }\eqno
(5) 
$$
and 
$$
\mathbf{A}(\mathbf{r},t)=\frac{\mu _{0}}{4\pi }\int \frac{\mathbf{J}(\mathbf{%
r}^{\prime },t)}{R}dv^{\prime },\eqno
(6) 
$$
where $\mu _{0}\epsilon _{0}=1/c^{2}$, $R=|\mathbf{r}-\mathbf{r}^{\prime }|$%
, and the time retardation $R/c$ from the source point $\mathbf{r}^{\prime }$
to the field point $\mathbf{r}$ is neglected. It is noted that compared to
the electrostatic force due to the scalar potential, both the electric
induction force and the magnetic force due to the vector potential are of
the second order of normalized speed with respect to $c$.

In railgun accelerators, the current $I$ flowing on the loop formed by the
rails, the armature, and the breech generates a magnetic vector potential $%
\mathbf{A}$ and a magnetic field $\mathbf{B}$. Then the current-carrying
armature experiences a magnetic force, which tends to accelerate the
armature to move along the rails. Correspondingly, there is another magnetic
force exerted on the breech as a recoil force. Meanwhile, the motion of the
armature results in another magnetic force on the armature itself. This
force is along the armature and then will counteract the electrostatic force
which in turn is established by an external power supply to support the
current $I$. The current depends on the resultant force and hence on the
speed of the armature. If the applied voltage is fixed, the current and
hence the magnetic vector potential will decrease. According to the Lorentz
force law, a time-varying vector potential will generate an electric
induction force. The electric induction force exerted on the ions of a
straight metal wire carrying a current decreasing with time is parallel to
the current. Thus the net induction force exerted on each rail of a railgun
will have a major component longitudinal to the rails. This force is not
expected to depend significantly on the location along each rail, while the
forces exerted on the respective rails are in opposite directions. As the
electric induction force is proportional to the time rate of change of the
current $I$, it depends on the acceleration of the armature.$\vspace{0.3cm}$

$\hspace{2.4cm}$\includegraphics[bb=0 0 4in 3.333in]{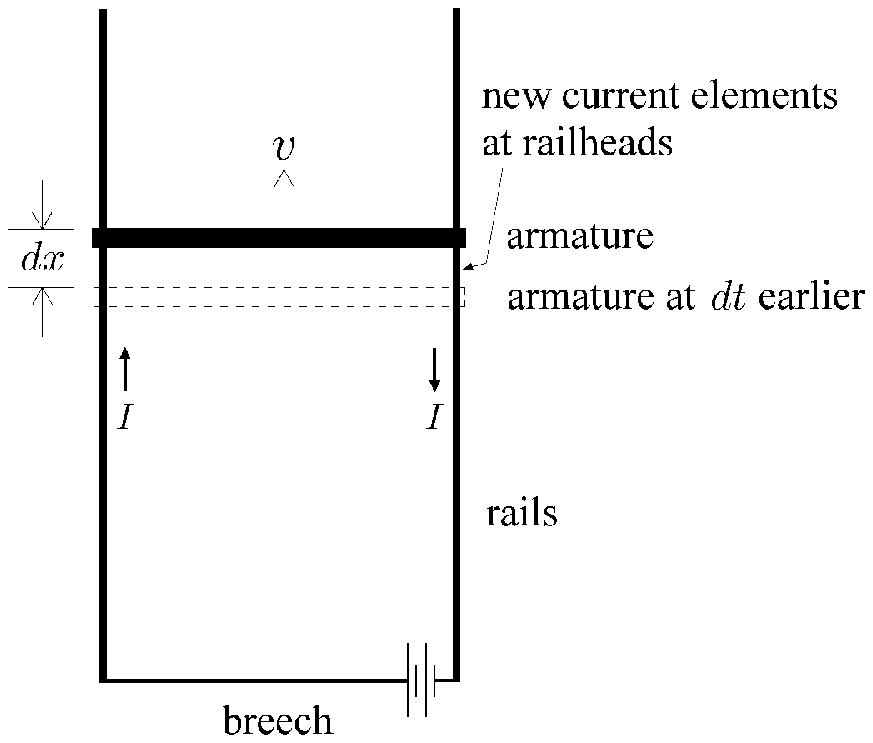}$\vspace{%
0.3cm}$

$\hspace{1.1cm}$%
\parbox{5in} {\baselineskip 0.1667in {
\QTR{bf}{Fig.\hspace{0.1cm}1}\quad The railgun accelerator. The movement of the armature along the rails constantly introduces new current elements which in turn generate the electric induction force on the rails.
}}\vspace{0.6cm}

Another effect of the motion of the armature is to constantly introduce new
current elements located on the rails just behind the armature, where the
current changes abruptly from zero to $I$, as depicted in Fig.\hspace{0.1cm}%
1. Accordingly, the magnetic vector potential has a tendency to increase
with time. (On the other hand, the current $I$ itself and hence the vector
potential tend to decrease as discussed previously.) This increment of the
vector potential is longitudinal to the rails and hence another electric
induction force longitudinal to them is induced. The vector potential due to
the new current elements is given by superposition. As the currents flowing
on the two rail segments of length $dx$ are in opposite directions, the
increment is given quantitatively by the difference 
$$
dA=\frac{\mu _{0}}{4\pi }\left( \frac{1}{x}-\frac{1}{\sqrt{x^{2}+s^{2}}}%
\right) Idx,\eqno
(7) 
$$
where $x$ is the distance of the observation position on one rail from the
moving armature and $s$ is the separation distance between the two rails. In
the preceding formula the cross section of the rail is supposed to be
vanishing; otherwise, the potential should be evaluated by a surface
integral over the cross section to get a more accurate result for a small $x$
and to avoid the singularity for a vanishing $x$. The length $dx$ introduced
during the movement of the armature over a short time interval $dt$ is
simply given by $vdt$, where $v$ is the speed of the armature with respect
to the rails. Thus the corresponding induction force exerted on an ion of
the rail is given by 
$$
F=-q\frac{dA}{dt}=-q\frac{\mu _{0}}{4\pi }\left( \frac{1}{x}-\frac{1}{\sqrt{%
x^{2}+s^{2}}}\right) Iv,\eqno
(8) 
$$
where the force is along the rail and $q$ is the charge of the ion. It is
noted that the force is proportional to the speed of the armature and the
current. These dependences are similar to those for the magnetic force. Thus
the induction force is of the same order of the magnetic force in magnitude.
Obviously, this electric induction force is the strongest at the
instantaneous heads of the rails. This situation agrees with the
experimental observation that the railheads were distorted significantly
after the launch of the armature \cite{Graneau87}.

Thus, in railgun accelerators, there are at least two electric induction
forces which are longitudinal to the rails and depend both on the speed and
on the acceleration of the armature. This force can also depend on the
location of the armature along the rails, as it determines the perimeter and
the resistance of the loop. Obviously, the electric induction force vanishes
for a substantially stationary armature, which is in agreement with some
similar experiments where it is found that the measured force is identical
to the calculated magnetic force \cite{Peoglos, Cavalleri98}. As the
aforementioned induction forces in railguns are in opposite directions, the
resultant induction force exerted on the ions of one rail can be parallel or
antiparallel to the direction of the current. In either case, the induction
forces on the respective rails are different in direction, if a direct
current is used in stead of an alternating current. This situation seems not
yet observed experimentally and deserves further investigation. Anyway, the
electric induction force should not be ignored in analyzing the longitudinal
recoil force in railguns.

Another mechanism for the recoil force may be the electrostatic force due to
internal sources, which is also ignored in the literature. The electrostatic
force is due to charges, stationary or moving, and can be much stronger than
the magnetic and induction forces by a factor like $(v/c)^{-2}$. In the
previous discussion of the induction force and the magnetic force,
electrical neutralization is assumed tacitly. If the neutralization is not
complete, a net electrostatic force will emerge and can dominate over the
other forces. According to the continuity equation, electric charges tend to
accumulate at the location where the current is not uniform, such as the
junctions between the rails and the armature and the interface between two
conductors of different conductivities. The electrostatic force may be used
to account for the experiment of the repulsion between a suspended $\pi $%
-shaped aluminum wire and the current-supplying wires, where the ends of the
wires are connected to mercury troughs. In this experiment it was observed
that the direction of the repulsion depends on the direction of the
current-supplying wires \cite{Pappa83}. Further, the wire fragmentation,
where a metal wire was observed to break into several segments after a high
current passed through it \cite{Graneau87b}, could be ascribed to a
complicated process involving a strong electrostatic force. However,
quantitative discussions of these electrostatic forces are difficult.

\vspace{1cm}

\noindent {\large \textbf{3. Derivation of Lorentz Force Law}}\\[0.2cm]
In classical mechanics the force exerted on a particle due to a potential
energy $U$ depending on the particle velocity $\mathbf{v}$ is given by
Lagrange's equation 
$$
\mathbf{F}=-\nabla U+\sum\limits_{i}\hat{i}\frac{d}{dt}\left( \frac{\partial
U}{\partial v_{i}}\right) ,\eqno
(9) 
$$
where $v_{i}=\mathbf{v\cdot }\hat{i}$, $\hat{i}$ is a unit vector, and the
index $i=x,y,z$.

It is known that the Lorentz force law (4) can be derived from Lagrange's
equation by adopting the velocity-dependent potential energy $U$ which in
turn incorporates the scalar potential $\Phi $ and the vector potential $%
\mathbf{A}$. That is, 
$$
U=q\Phi -q\mathbf{v\cdot A.}\eqno
(10) 
$$
This approach was pioneered by Clausius in 1877 \cite{ORahilly, Whittaker}.
In the derivation the expansion $d\mathbf{A}/dt=\partial \mathbf{A}/\partial
t+(\mathbf{v\cdot }\nabla )\mathbf{A}$ and the identity $\nabla (\mathbf{%
v\cdot A})-(\mathbf{v\cdot }\nabla )\mathbf{A=v}\times \nabla \times \mathbf{%
A}$ have been used. It is seen that the electric induction force is similar
to the magnetic induction force in their physical origin, where the latter
is associated with the term $(\mathbf{v\cdot }\nabla )\mathbf{A}$ and is an
ingredient of the magnetic force. In the preceding potential energy $U$, the
velocity $\mathbf{v}$ and the velocity of the mobile charged particles
involved in the potential $\mathbf{A}$ are not relative. Thus the potential
energy and hence the derived force are not frame-invariant under Galilean
transformations. Furthermore, the derived force between two moving charged
particles is not in accord with Newton's law of action and reaction. On the
other hand, it is known that the Lorentz force law is invariant under the
Lorentz transformation.

\vspace{1cm}

\noindent {\large \textbf{4. Weber Force Law and Amp\`{e}re Force Law}}\\[%
0.2cm]
In as early as 1846, Weber presented a second-order generalization of
Coulomb's law for electrostatic force. The Weber force law can be derived
from a velocity-dependent potential energy which, for the force exerted on a
particle of charge $q_{1}$ and velocity $\mathbf{v}_{1}$ due to another
particle of charge $q_{2}$ and velocity $\mathbf{v}_{2}$, is given by \cite
{ORahilly, Whittaker} 
$$
U=\frac{q_{1}q_{2}}{4\pi \epsilon _{0}}\frac{1}{R}\left( 1+\frac{u_{12}^{2}}{%
2c^{2}}\right) ,\eqno
(11) 
$$
where $R$ is the relative distance between the two charged particles, $%
u_{12}=(\mathbf{v}_{1}-\mathbf{v}_{2})\mathbf{\cdot }\hat{R}$ is the radial
relative speed between them, and $\hat{R}$ points from particle 2 to
particle 1.

As the potential energy depends on the radial speed, it is of convenience to
use the chain rule to express Lagrange's equation in the form 
$$
\mathbf{F}=-\nabla U+\sum\limits_{i}\hat{i}\frac{d}{dt}\left( \hat{i}\mathbf{%
\cdot }\hat{R}\frac{\partial U}{\partial u_{1}}\right) ,\eqno
(12) 
$$
where $v_{i}$ in (9) is understood as $v_{1i}$ ($=\mathbf{v}_{1}\mathbf{%
\cdot }\hat{i}$). Then, by using the identity $\nabla u_{12}=d\hat{R}/dt=(%
\mathbf{v}_{12}-u_{12}\hat{R})/R$, the preceding force formula becomes the
form given in \cite{Whittaker} 
$$
\mathbf{F}=\hat{R}\frac{1}{R}U+\hat{R}\frac{d}{dt}\frac{\partial U}{\partial
u_{1}}.\eqno
(13) 
$$
In dealing with the time derivative associated with the potential energy,
one uses the expansion $d(u_{12}/R)/dt=(du_{12}/dt)/R-u_{12}^{2}/R^{2}$, as
both of the variations of $u_{12}$ and $R$ contribute to the time
derivative. Further, by expanding the derivative $du_{12}/dt$, one has the
Weber force law \cite{ORahilly, Whittaker} 
$$
\mathbf{F}=\frac{q_{1}q_{2}}{4\pi \epsilon _{0}}\frac{\hat{R}}{R^{2}}\left(
1+\frac{v_{12}^{2}}{c^{2}}-\frac{3}{2}\frac{u_{12}^{2}}{c^{2}}+\frac{\mathbf{%
R}\cdot \mathbf{a}_{12}}{c^{2}}\right) ,\eqno
(14) 
$$
where $\mathbf{a}_{12}$ denotes the relative acceleration. It is noted that
the force is always along the radial direction represented by $\hat{R}$ and
the involved distance, velocity, and acceleration are all relative between
the two particles. Thereby, the Weber force is frame-invariant simply under
Galilean transformations and is in accord with Newton's law of action and
reaction.

Consider the case where the magnetic force is due to a neutralized current
where the mobile charged particles forming the current is actually embedded
in a matrix, such as electrons in a metal wire. The ions that constitute the
matrix tend to electrically neutralize the mobile particles. Suppose that
the various ions and hence the neutralizing matrix move at a fixed velocity $%
\mathbf{v}_{m}$. Thus the mobile charged particles drift at the speed $%
v_{2m} $ relative to the matrix. Ordinarily, the drift speed $v_{2m}$ is
quite low due to the collision of electrons against ions. Thus, based on the
Weber force law, the force due to a neutralized current element exerted on a
charged particle of relative velocity $\mathbf{v}_{1m}$ can be given by
superposing the forces due to the electron and ion. Thus one has the force
law between the current element and the particle 
$$
\mathbf{F}=\frac{q_{1}q_{2}}{4\pi \epsilon _{0}c^{2}}\frac{\hat{R}}{R^{2}}%
\left( -2\mathbf{v}_{1m}\cdot \mathbf{v}_{2m}+3u_{1m}u_{2m}-\mathbf{R}\cdot 
\mathbf{a}_{2m}\right) ,\eqno
(15) 
$$
where it has been supposed that the drift speed $v_{2m}$ is sufficiently low
as it is ordinarily and thus those terms associated with the second order of 
$\mathbf{v}_{2m}$ are neglected. It is noted that the term with $\mathbf{a}%
_{2m}$ is along the direction of $\hat{R}$, instead of the direction of $%
\mathbf{a}_{2m}$ itself. Consequently, the Weber force law disagrees with
the Lorentz force law as far as the longitudinal force in railgun
accelerators is concerned.

Consider two neutralized current elements flowing on two wire segments which
in turn are stationary with respect to each other. Then, by superposing the
forces exerted on the electron and ion, one has the force law between the
two current elements 
$$
\mathbf{F}=\frac{q_{1}q_{2}}{4\pi \epsilon _{0}c^{2}}\frac{\hat{R}}{R^{2}}%
\left( -2\mathbf{v}_{1m}\cdot \mathbf{v}_{2m}+3u_{1m}u_{2m}\right) .\eqno
(16) 
$$
This formula is identical to the Amp\`{e}re force law (2), as $q_{1}\mathbf{v%
}_{1m}$ and $q_{2}\mathbf{v}_{2m}$ correspond to $I_{1}d\mathbf{l}_{1}$ and $%
I_{2}d\mathbf{l}_{2}$, respectively. Since $\mathbf{v}_{1m}$ and $\mathbf{v}%
_{2m}$ are relative, the Amp\`{e}re force law is Galilean invariant. And as
these velocities appear in a symmetric way, the action of a current element
on itself then cancels out.

\vspace{1cm}

\noindent {\large \textbf{5. Riemann Force Law}}\\[0.2cm]
The electromagnetic force law can be derived alternatively from a potential
energy incorporating the relative speed, instead of the radial relative
speed. That is, \cite{ORahilly} 
$$
U=\frac{q_{1}q_{2}}{4\pi \epsilon _{0}}\frac{1}{R}\left( 1+\frac{v_{12}^{2}}{%
2c^{2}}\right) .\eqno
(17) 
$$
This velocity-dependent potential energy was introduced by Riemann in 1861 
\cite{Whittaker} and is almost ignored at the present time. Then Lagrange's
equation immediately leads to the Riemann force law \cite{ORahilly} 
$$
\mathbf{F}=\frac{q_{1}q_{2}}{4\pi \epsilon _{0}}\left\{ \frac{\hat{R}}{R^{2}}%
\left( 1+\frac{v_{12}^{2}}{2c^{2}}\right) -\frac{1}{c^{2}R^{2}}u_{12}\mathbf{%
v}_{12}+\frac{1}{c^{2}R}\mathbf{a}_{12}\right\} ,\eqno
(18) 
$$
where, as in deriving (14), one uses the expansion 
$$
\frac{d}{dt}\frac{\mathbf{v}_{12}}{R}=\frac{\mathbf{a}_{12}}{R}-\frac{u_{12}%
\mathbf{v}_{12}}{R^{2}},\eqno
(19) 
$$
as both of the variations of $\mathbf{v}_{12}$ and $R$ contribute to the
time derivative. Physically, the derivative $d(\mathbf{v}_{12}/R)/dt$ is
associated with the time rate of change in the potential energy actually
experienced by the affected particle. And the term with $u_{12}\mathbf{v}%
_{12}$ in the preceding force formula is associated with the variation of
the experienced potential energy due to the relative displacement between
the affected and the source particle. It is of essence to note that the
potential energy and the force depend on the relative velocity and distance
and hence they are independent of the choice of reference frames in uniform
motion of translation. Furthermore, the Riemann force law as well as the
Weber force law is in accord with Newton's third law of motion.

Now we consider the ordinary case where the force is due to a neutralized
current element with a sufficiently low drift speed $v_{2m}$. By
superposition the Riemann force exerted on a charged particle moving at a
velocity $\mathbf{v}_{1m}$ relative to the matrix is then given by 
$$
\mathbf{F}=\frac{q_{1}q_{2}}{4\pi \epsilon _{0}c^{2}}\left\{ \frac{1}{R^{2}}%
(-\hat{R}\mathbf{v}_{1m}\cdot \mathbf{v}_{2m}+u_{1m}\mathbf{v}_{2m}+u_{2m}%
\mathbf{v}_{1m})-\frac{1}{R}\mathbf{a}_{2m}\right\} .\eqno
(20) 
$$
Omitting the acceleration term, a similar force formula between two current
elements can be found in \cite{Whittaker}. When the current-carrying wire
forms a loop $C_{2}$ over which the current is uniform and thus the
neutralization remains, the force becomes 
$$
\mathbf{F}=\frac{q_{1}}{4\pi \epsilon _{0}c^{2}}\oint_{C_{2}}\frac{\rho _{l}%
}{R^{2}}(-\hat{R}\mathbf{v}_{1m}\cdot \mathbf{v}_{2m}+u_{1m}\mathbf{v}%
_{2m})dl\ -\ q_{1}\dfrac{\partial \mathbf{A}}{\partial t},\eqno
(21) 
$$
where $\rho _{l}$ denotes the line charge density of the mobile particles of
the neutralized loop, the vector potential is given by 
$$
\mathbf{A}=\frac{\mu _{0}}{4\pi }\oint_{C_{2}}\frac{\rho _{l}\mathbf{v}_{2m}%
}{R}dl,\eqno
(22) 
$$
and we have made use of the consequence that a uniform current ($\rho
_{l}v_{2m}$) leads to 
$$
\oint_{C_{2}}\frac{\rho _{l}u_{2m}}{R^{2}}dl=0.\eqno
(23) 
$$
Similarly, for a volume current density under the magnetostatic condition,
it can be shown that the contribution corresponding to that of the term $%
\rho _{l}u_{2m}$ cancels out collectively. It is noted that the time
derivative $\partial \mathbf{A/}\partial t$ is actually referred to the
matrix frame (in which the matrix is stationary) so that the variation of $%
\mathbf{v}_{2m}$ contributes to this derivative, while the variation of $R$
does not as its effect has been counted in the term with $u_{1m}\mathbf{v}%
_{2m}$ in (21).

Further, by using vector identities, the force given by (21) can be written
as 
$$
\mathbf{F}=q_{1}\left\{ \mathbf{v}_{1m}\times (\nabla \times \mathbf{A)}-%
\dfrac{\partial \mathbf{A}}{\partial t}\right\} .\eqno
(24) 
$$
It is of essence to note that the preceding formula looks like the Lorentz
force law under neutralization. However, the current density generating the
potential $\mathbf{A}$, the time derivative of $\mathbf{A}$ in the electric
induction force, and the particle velocity connecting to $\nabla \times 
\mathbf{A}$ in the magnetic force are all referred specifically to the
matrix frame. It is noted that this specific frame has been adopted tacitly
in common practice with the magnetic and induction forces. Thus, for the
force due to closed circuits, the Riemann force law which is Galilean
invariant and in accord with Newton's law of action and reaction can be
identical to the Lorentz force law. Recently, based on a wave equation a
time evolution equation similar to Schr\"{o}\-dinger's equation is derived.
From the evolution equation an electromagnetic force given in a form quite
similar to Lagrange's equation in conjunction with the potential energy (17)
is derived \cite{Su, QEM}. Thus a quantum-mechanical basis for the Riemann
force law has been provided. Further, the divergence and the curl relations
for the corresponding electric and magnetic fields are derived. Under the
magnetostatic condition, these four relationships are just Maxwell's
equations, with the exception that the velocity determining the involved
current density is also relative to the matrix \cite{QEM}.

\vspace{1cm}

\noindent {\large \textbf{6. Conclusion}}\\[0.2cm]
It is shown that in railgun accelerators the electric induction force
longitudinal to the rails is generated during the movement of the armature.
This force is due to the decrease of the current and to the newly introduced
current elements. Thus it depends on the location, speed, and acceleration
of the armature. This induction force is comparable to the magnetic force in
magnitude and has a tendency to be the strongest at the railheads. Thus it
accounts for the observed longitudinal recoil force exerted on the rails.
Besides, we compare the Weber and the Riemann force law, which are derived
from Lagrange's equation in conjunction with a potential energy depending on
the radial relative speed and on the relative speed, respectively. For
ordinary cases where the force is due to the current on a neutralized and
closed wire with low drift speed, it is shown that the Riemann force law
reduces to the Lorentz force law. Thus the longitudinal force exerted on the
rails of railgun accelerators can be well accounted for by the Riemann force
law which is in accord both with the nowadays standard theory and with
Newton's law of action and reaction.


\vspace{1cm}

\begin{thebibliography}{99}
\bibitem{Graneau87}  P. Graneau, \textit{J. Appl. Phys}. \textbf{62}, 3006
(1987).

\bibitem{Jolly}  D.C. Jolly, \textit{Phys. Lett. A} \textbf{107}, 231 (1985).

\bibitem{Ternan}  J.G. Ternan, \textit{J. Appl. Phys.} \textbf{57}, 1743
(1985).

\bibitem{Moyssides}  P.G. Moyssides, \textit{IEEE Trans. Magn.} \textbf{25},
4307 (1989).

\bibitem{Assis96}  A.K.T. Assis and M.A. Bueno, \textit{IEEE Trans. Magn.} 
\textbf{32}, 431 (1996).

\bibitem{Graneau01}  P. Graneau and N. Graneau, \textit{Phys. Rev. E}. 
\textbf{63}, 058601 (2001).

\bibitem{Cavalleri01}  G. Cavalleri, E. Tonni, and G. Spavieri, \textit{%
Phys. Rev. E} \textbf{63}, 058602 (2001).

\bibitem{Pappa83}  P.T. Pappas, \textit{Nuovo Cimento B} \textbf{76}, 189
(1983).

\bibitem{Phipps}  T.E. Phipps and T.E. Phipps Jr., \textit{Phys. Lett. A} 
\textbf{146}, 6 (1990).

\bibitem{Peoglos}  V. Peoglos, \textit{J. Phys. D} \textbf{21}, 1055 (1988).

\bibitem{Cavalleri98}  G. Cavalleri, G. Bettoni, E. Tonni, and G. Spavieri, 
\textit{Phys. Rev. E} \textbf{58}, 2505 (1998).

\bibitem{Graneau87b}  P. Graneau, \textit{Phys. Lett. A} \textbf{120}, 77
(1987).

\bibitem{ORahilly}  A. O'Rahilly, \textit{Electromagnetic Theory} (Dover,
New York, 1965), vol. 1, ch. 7; vol. 2, ch. 11.

\bibitem{Whittaker}  E. T. Whittaker, \textit{A History of the Theories of
Aether and Electricity} (Amer. Inst. Phys., New York, 1987), vol. 1, chs. 3,
7, and 13.

\bibitem{Su}  C.C. Su, \textit{J. Electromagnetic Waves Applicat.} \textbf{16%
}, 1275 (2002).

\bibitem{QEM}  C.C. Su, \textit{Quantum Electromagnetics -- A Local-Ether
Wave Equation Unifying Quantum Mechanics, Electromagnetics, and Gravitation}
(\texttt{http://qem.ee.nthu.edu.tw}).
\end{thebibliography}
\end{document}